\documentclass[prd,one column,12pt]{revtex4-1}
\usepackage{graphicx}
\usepackage[mathlines]{lineno}
\usepackage{amsmath}

\usepackage{slashed}
\usepackage{natbib}
\usepackage{bm}
\usepackage{amsfonts}
\usepackage{amssymb}
\usepackage{booktabs}
\usepackage{epsfig}
\usepackage{subfigure}
\usepackage{hyperref}
\usepackage{soul}
\usepackage{cancel}
\usepackage{ulem}
\usepackage{xcolor}
\usepackage{color}
\usepackage[toc,page]{appendix}

 \setul{0.5ex}{0.3ex}
\definecolor{Blue}{rgb}{0.0,01.0,1.5}

\begin{document}
	\newcommand\bluesout{\bgroup\markoverwith{\textcolor{blue}{\rule[0.5ex]{2pt}{0.4pt}}}\ULon}
\title{\textbf{Neutrino induced vorticity, Alfv\'en waves and the normal modes}}

\author{Jitesh R. Bhatt$^{1}$}
 \email[e-mail: ]{jeet@prl.res.in}
\author{Manu George$^{1, 2}$}
\email[e-mail: ]{manu@prl.res.in}
\affiliation{%
\centerline{$^{1}$ Physical Research Laboratory, Theory Division, Ahmedabad 380 009, India}\\
\centerline{$^{2}$ Department of Physics, Indian Institute of Technology, Gandhinagar, Ahmedabad 382 424, India}
}

\date{\today}
 \bigskip
 

\begin{abstract}
We consider  plasma consisting of electrons and ions in presence of a background  neutrino gas and develop the magnetohydrodynamic equations for the  system. We show that electron neutrino interaction can induce vorticity in the plasma even in the absence of any electromagnetic perturbations if the background neutrino density is  left-right asymmetric. This induced vorticity support a new kind of Alfv\'en wave whose velocity depends on both the external magnetic field and on the neutrino asymmetry. The normal mode analysis show that in the presence of neutrino background the Alfv\'en waves can have different velocities. We also discuss our results in the context of dense astrophysical plasma such as magnetars and  show that the  difference in the Alfv\'en velocities  can be used to explain the observed pulsar kick.  We discuss also the relativistic generalization of electron fluid in presence of asymmetric neutrino background.

\end{abstract}

\maketitle


\section{Introduction}
\label{intro}

It is important to study the characteristics  of plasma in presence of neutrinos, since such systems are important in understanding various physical phenomena during  the evolution of early Universe as well as the systems like core core-collapsing supernovae and magnetars (see for e.g. \cite{volpe_14} for a brief overview). The presence of the cosmic neutrino background can influence cosmic microwave anisotropy  and matter clustering \cite{komatsu_10,bashinsky_04} and it can also influence  dynamics of the primordial magnetic field \cite{kandu_98,jedamzik_98,shaw_10}. There exists several studies in literature where the neutrino plasma interaction has been analysed in a variety of physical situations. Non-linear coupling of intense neutrino flux with collective plasma oscillations is studied in the Ref. \cite{Bingham_94}. Authors  have shown that a neutrino flux as intense as that in supernovae core can cause parametric instabilities in the surrounding plasma. Effect of a neutrino medium in the evolution of lepton plasma had been  studied invoking ponderomotive description \cite{Shukla_1997,Bethe_96}. In these cases it was shown that the ponderomotive force is proportional to the gradient of neutrino density and   the electrons are repelled from the regions where neutrino density is large. Interaction of very large number of neutrinos with collective plasma and oscillation and the excitation of plasma turbulence is considered in the Ref. \cite{Tsytovich_98}. Different kinds of plasma neutrino interactions using the ponderomotive force  description and the effect on collective plasma properties can be found in the references, \cite{Silva_99,Bento_99,Esposito_99,Brizard_00,Bento_01,Serbeto_02,Serbeto_fluid_02}. In the fore mentioned ponderomotive force description, it was assumed that the neutrino field satisfy the naive  Klein-Gordon equation with appropriate interaction terms. Thus in this formalism the information about the chiral structure of the weak interaction is absent. Here we note that by Silva {et. al} in Refs. \cite{Silva_1999} and \cite{Silva_2006} the problem of neutrino driven streaming instability, which in turn can generate an significant energy transfer from neutrino to the plasma, was considered in the kinetic theory formalism. 
Formulation to study the plasma interaction with intense neutrino beam using the field theory techniques  is developed in \cite{Bento_99_recent}. Photon polarization tensor in a medium consistent with gauge and Lorentz invariance can be found in \cite{Palash_89}. In this work it is shown that, in presence of a medium, the photon polarization tensor can have  anti-symmetric part indicating $P$ and $CP$ violations. Further studies of such effect in presence of neutrinos for different physical scenarios are explored in [\cite{Nieves_00},\cite{Nieves_05}].

 In the context of early Universe, it has been shown by Shukla {\it at. el.} \cite{Shukla_1997}that, the ponderomotive force of non-uniform intense neutrino beam  can be responsible for large scale quasi-stationary magnetic field. In fact, later was the first one to suggest magnetic field generation in plasma due to plasma-neutrino interactions.  Further, large-scale magnetic field generation at  the time of neutrino decoupling due to the evolution of   plasma in presence of asymmetric neutrino background  is studied in \cite{Boyarsky_12} and \cite{Doglov_02}. This field  can act as a seed  for generation of the galactic magnetic field via the galactic  dynamo mechanism (see e.g \cite{Vainshtein_72} to read about galactic dynamo mechanism). It is to be noted that at finite lepton/baryon density the loop corrections to the photon polarization tensor are non-vanishing. 
With these corrections the photon polarization tensor acquires a non-zero parity odd contribution $\Pi_2(\bm k)$ where, $\bm k$ is the wave vector.
A finite and non-zero values of $\Pi_2(\bm k)$ in the photon polarization tensor means that there can be single field derivative terms in the effective Lagrangian and free energy, which dominates the kinetic energy part of the free energy which is having double derivative term. For e.g. the free energy for a static gauge field can be written as $\mathcal{F}[A]=\int d^3pA_i(k)\Pi_{ij}(k)A_j(-k)$ and with parity violating interactions $\Pi_{ij}(k)$ can have a contribution $i\Pi_2(p^2)\epsilon_{ijl}k_l$. Thus a non-zero value of $\Pi_2(0)$ means a term $\Pi_2(0)\bm A\cdot\nabla\times\bm A$ in the expression for free energy.
This in turn means that there can be a generation of a large scale ($k\rightarrow 0$) magnetic field by an instability arising due to  non-zero values of parity-odd contributions $\Pi_2(0)$ to the polarization tensor \cite{Boyarsky_12}. In the ref. \cite{Dornikov_15}, thermal field theory calculations were carried out to study the corrections to the photon polarization in presence of a background neutrino which is asymmetric in left-right number densities. Authors have shown that the axial part $\Pi_{2}$ is proportional to the neutrino asymmetry parameter and argued that  the contribution to $\Pi_{2}$ due to the plasma which is interacting with the neutrino gas is $\sim 10^1$ times larger than the contribution to $\Pi_2$ through the correction due to the virtual process. In ref. \cite{Bhatt_16} using a kinetic theory approach
it was shown that the photon polarization tensor can have the parity-odd contribution 
$\Pi_2(\bm k)$  due to the asymmetric neutrino background in both the collision less and
collision dominated regime. In the collision dominated regime the result for $\Pi_2(\bm k)$
using the kinetic approach agrees with that in ref. \cite{Dornikov_15}.
 In a recent work \cite{Diaz_16} authors have calculated the effective potential or refractive index for the cosmic neutrino background (CNB) and future experimental implications have been discussed.

Further, recent theoretical calculations showed that the asymmetry in the neutrino density can be transmuted to the fluid helicity  for sufficiently large electron neutrino interaction\cite{Yamamotto_16}. This neutrino induced   vorticity can act as axial chemical potential for the chiral electrons. This phenomenon can induce {\it helical plasma instability} that generate strong magnetic field \cite{Yamamotto_16}.  In this work the plasma particles
are considered to be massless and chirally polarized. Moreover, it was assumed, in this work, that the neutrino mean free path $l_\nu$ is much smaller than the system dynamics
at the length scale $L$ i.e. $L\gg l_\nu$. This allows one to write the equations for the neutrino
hydrodynamics \cite{Yamamotto_16}. 
Though this assumption is justifiable for a core collapsing supernova, it is hard to be satisfied in other scenarios like the early Universe. Electroweak plasma in a rotating matter is studied in \cite{Dornikov_15}. In this work it is shown that electric current can be induced in the direction of rotation axis due to the parity violating nature of the interaction. This phenomenon is called \textit{galvano-rotational effect} (GRE). In a recent work \cite{Suvorov_16},  spin paramagnetic deformation of neutron star has been studied and authors have calculated the ellipticity of a strongly magnetized neutron star using the spin magneto hydrodynamic  equations developed in \cite{Brodin_07}.


In the present work we are interested in developing magnetohydrodynamic description of the plasma
in presence of the left-right asymmetric neutrino background.  The expression of the interaction Lagrangian of a charged lepton field and the asymmetric neutrinos suggests that the neutrino can couple with spin of the electron [\cite{Giunti_07},\cite{Dornikov_15}]. It is interesting to note here that there exist a lot of literatures in the usual electron-ion plasmas where the dynamics of  spin
degree  can play a significant role. For example it was suggested that a spin polarized
plasma in a fusion reactor can yield higher nuclear reaction cross section \cite{Kulsrud_82} and the spin depolarization process in the plasma can remain small \cite{Cowley_86}. Effect of  spin dynamics  using single particle description, valid for  a dilute gas, is studied   in the context of laser plasma interaction in the Ref. \cite{Walser_02}. 
The collective effects within the framework of spin-magnetohydrodynamics has been studied
in Refs.\cite{Brodin_07,Mahajan_11} ( for general discussion see \cite{Haas_11}). These works  can have applications in studying environments with a strong external magnetic field like pulsar and magnetars. In the present work we generalize the spin-magnetohydrodynamics considered in \cite{Brodin_07}
to incorporate the effect of asymmetric neutrino background. 

The report is organised in the following way. In section \ref{lag_non_rel} we consider the low energy Lagrangian  for our system and the equations of motion and spin evolution equations are derived invoking the non-relativistic  approximations. MHD equations are considered in section \ref{hydro_eqns}. Velocity perturbations and electromagnetic perturbations in a magnetized  plasma interacting with neutrino background  is considered in section \ref{mode_analysis}.In section \ref{pul_kick} we apply our theory to neutron star to calculate the kick  and section \ref{discussion} is about summary and conclusions. We provide a brief summery of relativistic generalization of the theory in appendix \ref{app_A}.


\section{The Lagrangian and non-relativistic approximation}
\label{lag_non_rel}
Lagrangian density for lepton field interacting with background neutrino is given by,
\begin{equation}
 \mathcal{L}=\bar\psi[i\gamma^\mu\partial_\mu\psi-\gamma_\mu(f^\mu_LP_L+f^\mu_RP_R)-m]\psi
 \label{lagrangian} 
\end{equation}
where, m is mass of the lepton, $\gamma^\mu=(\gamma^0, \bm{\gamma})$ are the Dirac matrices and $P_{L,R}=\frac{1\mp\gamma^5}{2}$ are the 
chiral projection operator with $\gamma^5=i\gamma^0\gamma^1\gamma^2\gamma^3 $.  $f^\mu_{L,R}=(f^0_{L,R}, \bm{f}_{L,R})$
are the neutrino currents and they are regarded as an external macroscopic quantities.

An explicit form of $f^\mu_{L,R}$ can be calculated from effective Lagrangian (\cite{Giunti_07}, \cite{Dornikov_15})

\begin{equation}
\mathcal{L}_{eff}=[-\sqrt{2} G_F\sum_\alpha\bar\nu_\alpha\gamma^\mu\frac{(1-\gamma^5)}{2}\nu_\alpha][\bar\psi\gamma_\mu(a_L^\alpha P_L+a_R^\alpha P_R)\psi]
\label{Leff}
\end{equation}

where, label $\alpha$ denotes neutrino species $\alpha=e,\,\mu,\,\tau$ and  $G_F=1.17\times10^{-11}MeV^{-2}$ is the Fermi constant. The coefficients $a_L^\alpha$ \& $a_R^\alpha$
are given by

\begin{equation}
 a_L^\alpha=\delta_{\alpha,e}+sin^2{\theta_W}-1/2 , a_R^\alpha=
 sin^2{\theta_W}
 \label{aLR}
\end{equation},

with $\theta_W$ being the Weinberg angle. Next, we assume that  $\nu \bar\nu$ form an isotropic background gas.
 This in turn means that in averaging over the
neutrino ensemble, only non-zero quantity will be\, $<\bar\nu_\alpha\gamma^0(1-\gamma^5)\nu>=2(n_{\nu_\alpha}-n_{\bar\nu_\alpha})$. Number densities 
of neutrinos and anti-neutrinos can be calculated using corresponding  Fermi-Dirac distribution function

\begin{equation}
 n_{\nu_\alpha , \bar\nu_\alpha}=\int \frac{d^3p}{(2\pi)^3}\frac{1}{e^{\beta_{\nu_\alpha}(|\bm p|\mp\mu_{\nu_\alpha} )}+1}
 \label{FDdistribution}
\end{equation}

where $\beta$ is the inverse temperature. Using Eqns. (\ref{lagrangian}-\ref{FDdistribution}) one obtains 

\begin{align}
  f^0_L=&2\sqrt 2 G_F[\Delta n_{\nu_e}+(sin^2\theta_W-1/2)\sum_\alpha\Delta n_{\nu_\alpha})] ,\\  f^0_R=&2\sqrt 2 G_F sin^2
 \theta_W\sum_\alpha\Delta n_{\nu_\alpha}.
\end{align}

Thus the equation of motion obtained from Eqn. (\ref{lagrangian}) can be written as 

\begin{equation}
 i\frac{\partial\psi}{\partial t}=[\bm\alpha\cdot\bm{\hat p}\psi+\beta m-(f^0_LP_L+f^0_RP_R)]\psi.
 \label{eompsi}
\end{equation}

Writing $\psi=\left(\begin {array}{c} \phi \\ \chi \end {array}\right ) $ in the Eqn.(\ref{eompsi}) and following the standard procedure \cite{Greiner_reqm}, Hamiltonian
for the  large component of the spinor  can be obtained as,

\begin{equation}
\mathcal{ H}=\frac{1}{2m}\bm{(\sigma\cdot p)(\sigma\cdot p)}+\frac{\Delta f^0}{2m}\bm{(\sigma\cdot p)}+\frac{f^0}{2}+O(f_{L,R}^2)
\nonumber
\end{equation}

where, $f^0=f^0_L+f^0_R$ and $\Delta f^0=f^0_L-f^0_R$. In the above equation, we have neglected terms proportional to $G_F^2$.
In the presence of external electromagnetic field,  momentum $\bm p$ has to be 
replaced by $\bm{p-eA}$. Thus the Hamiltonian for charged fermion in interacting with an external electromagnetic
field and background neutrino is given by,

\begin{equation}
  \mathcal {H}=
 \frac{(\bm{p-eA})^2}{2m}-\bm{\mu\cdot B}+eA^0+\frac{\Delta f^0}{2m}\bm\sigma\cdot(\bm p-e\bm A)+\frac{f^0}{2}
\label{H}
\end{equation}

\noindent
where, $\bm\mu=\frac{eg}{4m}\bm\sigma$ is the electron magnetic moment and $g$ is the Land\`{e} g-factor. The first three terms on the right hand
side  are well known and very well studied in the literature. The fourth and fifth terms are due to the neutrino background. The last term 
might contribute
to  the energy of the system, but it will not enter into the equations of motion as the neutrino background considered to be constant. 
If the neutrino background vary with space and time, this term would modify force 
equation as $\bm F\propto \bm\nabla f=\bm\nabla\psi_\nu^*\psi_\nu$. This force is called ponderomotive force. Such a scenario was studied 
in Ref. \cite{Bethe_96}, however in their formalism
the fourth term was not considered. 

 In order to find the equation of motion
for a charged particle in an electromagnetic
field and the neutrino background, one can
use Eqn.(\ref{H}) and the 
Heisenberg equation $\dot{\hat O}= i[\hat {\mathcal{H}},\hat O]$ and write: 
\begin{equation}
 \bm v=\dfrac{\bm p-e\bm A}{m}+\frac{\Delta f^0}{2m}\bm\sigma
 \label{dotx}
\end{equation}
where we wrote $\dot{\bm x}=\bm v$.

\begin{equation}
 \dot{\bm p}=\frac{e}{m}(\bm p-e\bm A)_k\bm\nabla\bm A_k+\frac{eg}{m}\bm\nabla\bm{(s\cdot B)}-e\bm{\nabla}A^0
 \label{dotp}
\end{equation}

where we have defined $\bm s=\bm\sigma /2$ and
\begin{equation}
 \dot{\bm s}=\mu_B(\bm s\times\bm B)-\Delta f^0(\bm s\times\bm v)
 \label{spin_dyn}
\end{equation}
From the equations \ref{dotx}-\ref{spin_dyn} we get 

\begin{equation}
 \ddot{\bm x}=\frac{e}{m}[\bm E+\bm v\times\bm B]+\frac{e\Delta f^0}{2m^2}(\bm s\times \bm B)+\frac{eg}{2m^2}\bm\nabla(\bm s\cdot\bm B)
 \label{ddotx}
\end{equation}


\section{The hydrodynamic equations}
\label{hydro_eqns}
In this section we follow the methods developed in Ref. \cite{Brodin_07} to derive the hydrodynamic equations from the quantum Lagrangian for spin half particles. We consider a system of electrons and ions in presence of homogeneous neutrino background. The neutrino background assumed to have a left-right asymmetry. Furthermore we treat electrons as quantum particles and ions as classical particles so that we can neglect the spin dynamics and other quantum effects for ions.

 For simplicity, first let us Consider the  Eqn. (\ref{H}) without the neutrino interaction term. We can  decompose  the wave function as  $\psi_\alpha =\sqrt{n_\alpha} e^{iS_\alpha}\chi_\alpha$,  where $n_\alpha$ is the density, $S_\alpha$ is the phase and $\chi_\alpha$ is a two component spinor in which the spin-1/2 information is contained. Inserting this decomposition and considering the real and imaginary parts of the resulting equation we get the continuity and momentum conservation equation for the "species $\alpha$" as,

\begin{equation}
	\frac{\partial n_\alpha}{\partial t}+\nabla \cdot (n_\alpha \bm{\upsilon}_\alpha)=0
	\label{cont_spe}
\end{equation}

and

\begin{align}
	&\nonumber m_\alpha(\frac{\partial}{\partial t}+\bm\upsilon_\alpha\cdot\bm{\nabla})\bm\upsilon_\alpha  =
	  q_\alpha(\bm E+\bm\upsilon_\alpha\times\bm B) +\\ & 2\mu(\nabla\otimes\bm B)\cdot\bm s_\alpha-\nabla Q_\alpha-\frac{1}{m_\alpha n_\alpha}\nabla\cdot (n_\alpha
	\bm\Sigma_\alpha)
	\label{mom_cons_spe}
\end{align}

Velocity is defined via $m_\alpha\bm\upsilon_\alpha=\bm j_\alpha/\psi^\dagger\psi$, from which we obtain

\begin{equation}
m_\alpha\bm\upsilon_\alpha=(\nabla S-i\chi_\alpha^\dagger\nabla\chi_\alpha)-q_\alpha\bm A	
\label{vel_spe_def}
\end{equation}

and,

\begin{equation}
	\bm s_\alpha=\frac{1}{2}\chi_\alpha^\dagger\bm\sigma\chi_\alpha
\end{equation}

The quantity $Q_\alpha$ is known as the quantum potential(Bohm potential) defined as

\begin{equation}
	Q_\alpha=-\frac{1}{2m_\alpha\sqrt n_\alpha}\nabla^2\sqrt n_\alpha
\end{equation}

and $\bm \Sigma_\alpha$ is the symmetric spin gradient tensor.

\begin{equation}
	\bm \Sigma_\alpha=\nabla\bm s_{(\alpha)a}\otimes\nabla\bm s_{(\alpha)}^a
\end{equation}

where $a=1,2,3$. By contracting the Pauli equation with $\psi^\dagger\bm\sigma$, one can obtain the spin evolution equation
as

\begin{equation}
	(\frac{\partial}{\partial t}+\bm\upsilon_\alpha\cdot\bm{\nabla})\bm s_\alpha=2\mu (\bm s_\alpha\times\bm B)+\frac{\bm s_\alpha\times[\partial_a(n_\alpha\partial^a\bm s_\alpha)]}{m_\alpha n_\alpha}
	\label{spin_evo_spe}
\end{equation}

In presence of neutrino background, the continuity equation remain unchanged. But both momentum conservation  and spin evolution equations are modified in the following way.

\begin{align}
	&\nonumber m_\alpha(\frac{\partial}{\partial t}+\bm\upsilon_\alpha\cdot\bm{\nabla})\bm\upsilon_\alpha  =
	  q_\alpha(\bm E+\bm\upsilon_\alpha\times\bm B) +\\ & 2\mu(\nabla\otimes\bm B)\cdot\bm s_\alpha-\nabla Q_\alpha-\frac{1}{m_\alpha n_\alpha}\nabla\cdot (n_\alpha\bm\Sigma_\alpha)+\frac{\Delta f^0}{2m}\bm s_\alpha\times\bm B
	  \label{mom_cons_spe_nu}
\end{align}

and

\begin{equation}
	(\frac{\partial}{\partial t}+\bm\upsilon_\alpha\cdot\bm{\nabla})\bm s_\alpha=2\mu (\bm s_\alpha\times\bm B)-\frac{\Delta f^0}{2}\bm s_\alpha\times\bm\upsilon_\alpha+\frac{\bm s_\alpha\times[\partial_a(n_\alpha\partial^a\bm s_\alpha)]}{m_\alpha n_\alpha}
	\label{spin_evo_spe_nu}
\end{equation}

Now, in order to define hydrodynamic quantities, we need to specify how to calculate the expectation values. Suppose that we have N wave function with same kind of particles with magnetic moment $\mu$  charge $q$ and mass $m$ so that the wave function for the entire system can be factorised as $\psi=\psi_{(1)}\psi{(2)}...\psi_{(N)}$. Then we can define the total particle density for charge $q$ as, $n_q=\sum_{\alpha}n_\alpha$ and the expectation value of any quantity $f$ as $<f>=\sum_\alpha\frac{n_\alpha}{n_q}f$.

Using these arguments we define the total fluid velocity $\bm V_q=<\bm\upsilon_\alpha>$ and $\bm S_q=<\bm s_\alpha>$. In order to simplify further calculations, we redefine  these quantities such that $\bm w_\alpha=\bm \upsilon_\alpha-\bm V_q$ and $\bm{\mathcal S}_\alpha=\bm s_\alpha-\bm S_q$, satisfying $<\bm w_\alpha>=0$ and $<\bm{\mathcal S}_\alpha>=0$. Now taking the ensemble average of the equations (\ref{cont_spe}), (\ref{mom_cons_spe_nu}) and (\ref{spin_evo_spe_nu}) we get the following expressions. 
\begin{equation}
	\frac{\partial n_q}{\partial t}+\nabla\cdot(n_q\bm V_q)=0
\end{equation}
\begin{equation}
	m_q n_q\Big(\frac{\partial}{\partial t}+\bm V_q\cdot\nabla\Big)\bm V_q=qn_q\Big(\bm E+\bm V_q\times\bm B\Big)-\nabla\cdot\bm\Pi-\nabla P+\bm{\mathcal{C}}_{qi}+\bm F_Q+\bm F_{\nu e}
	\label{Hydro_eqn_V_q}
\end{equation}
and
\begin{equation}
	n_q\Big(\frac{\partial}{\partial t}+\bm V_q\cdot \nabla\Big)\bm S_q=2\mu_B n_q\bm S_q\times\bm B-\frac{\Delta f^0}{2}\bm S_q\times \bm V_q+\bm\Omega_s-\nabla\cdot \bm {K}_q+\bm K_{\nu e}
	\label{Hydro_eqn_S_q}
\end{equation}
Where, $\bm\Pi$ is the traceless anisotropic part of the pressure tensor and $P$ is the homogeneous part. $\mathcal{C}_{qi}$ represents the collision between particle with charge $q$ and ion denoted using the letter $i$ and the quantum force density $\bm F_Q$ and force due to the interaction with the neutrino back ground $\bm F_{\nu e}$ has the definitions,
\begin{align}
	\bm F_Q&=2\mu_B n_q\Big(\nabla\otimes\bm B\Big)\cdot\bm{S}_q-n_q\Big<\nabla Q_\alpha \Big>-\frac{1}{m}\nabla\cdot\Big(n_q\bm\Sigma\Big)-\frac{1}{m}\nabla\cdot\Big(n_q\tilde\Sigma\Big)	\label{F_Q}
 \\
	\nonumber &-\frac{1}{m}\nabla\cdot\Big[\Big(\nabla\bm S_a\Big)\otimes\Big<\nabla\mathcal{S}^{a}_\alpha\Big>+n_q\Big<\nabla\mathcal{S}^{a}_\alpha\Big>\otimes\Big(\nabla\bm S_a\Big)\Big]
\end{align}
and 
\begin{equation}
	\bm F_{\nu e}=n_ee\frac{\Delta f^0}{2m}\bf S_q\times \bm B
\end{equation}
The quantities $\bm \Omega_s, \bm\Sigma$ and $\bm{\tilde\Sigma}$ depends on the spin of the particles and their precise definitions can be found in the Ref. \cite{Brodin_07}. $\bm K_q=\Big<\mathcal{\bm S}_{\alpha }\otimes  \bm w_{\alpha }\Big>$ is the spin thermal coupling and $\bm K_{\nu e}=\epsilon_{ijk}\frac{\delta f^0}{2}\Big<\mathcal{\bm S}_{\alpha j}\bm w_{\alpha k}\Big >$ is the thermal spin coupling induced by neutrino interaction.

In the following sections, we will replace the subscript $q$ with $e$ and $i$ for electrons and ions respectively. Since we are considering ions as classical particle, we can neglect the contributions from spin and other quantum effects for ions. Thus, the fluid equations for ions read,
\begin{equation}
	\frac{\partial n_i}{\partial t}+\nabla\cdot(n_i\bm V_i)=0
	\label{cont_ion}
	\end{equation}
	\begin{equation}
	m_i n_i\Big(\frac{\partial}{\partial t}+\bm V_i\cdot\nabla\Big)\bm V_i=q_in_i\Big(\bm E+\bm V_i\times\bm B\Big)-\nabla\cdot\bm\Pi_i-\nabla P_i+\bm{\mathcal{C}}_{iq}
	\label{mom_ion}
\end{equation}
Note that there is no spin evolution equation for ions. Therefore whatever spin contributions  governing the dynamics of the system are only due to the spin of the electrons. Now we can construct the single fluid equations from the above equations for electrons and ions. In order to do that we define the total mass density, $\rho=(m_en_e+m_in_i)$, the centre of mass velocity of the fluid $\rho\bm V=(m_en_e \bm V_e+m_in_i\bm V_i)$    and the current density $\bm j=(-en_e\bm V_e+Zen_i\bm V_i)$ and assuming quasi-neutrality $n_e=Zn_i$, one can immediately obtain the continuity equation,	
\begin{equation}
	\frac{\partial\rho}{\partial t}+\nabla\cdot(\rho\bm V)=0
	\label{cont_flu}
\end{equation}

 and momentum conservation equation
 
 \begin{equation}
 	\rho\Big(\frac{\partial}{\partial t}+\bm V\cdot\nabla\Big)\bm V=\bm j\times\bm B-\nabla\cdot\bm\Pi-\nabla P+\bm{F}_Q+\bm{F}_{\nu e}
 \end{equation}
 
Note that, with the assumption of quasi-neutrality we can write $n_e=\rho /(me_e+m_i)$ and $\bm V_e=\bm V-m_i\bm j/Ze\rho$. There fore we can express the quantum terms in terms of the total density and , centre of mass velocity of the fluid and  current. Thus the spin evolution equation becomes,

\begin{align}
	\rho\Big(\frac{\partial}{\partial t}+\bm V\cdot\nabla\Big)\bm S&=\frac{m_i}{Ze\rho}\bm j\cdot\nabla \bm S+2\mu\rho\bm S\times\bm B-\Big(m_e+m_i/Z\Big)\nabla\cdot\bm K_e+\Big(m_e+m_i/Z\Big)\bm{\Omega}_s\\\nonumber &-\frac{e\Delta f^0\rho}{2m_e} \bm S\times\Big(\bm V-\frac{m_i\bm j}{Ze\rho}\Big)-\frac{e\Delta f^0}{2m_e}\Big<\bm w_e\times\mathcal{\bm S}_e\Big>
\end{align}

in general, for a magnetized medium with magnetization density $\bm M$ we can write the free current density $\bm j=\frac{\nabla\times\bm B}{\mu_0}-\bm j_M$, where $\bm {j}_M=\nabla\times \bm M$ is the magnetization current density. Note that, here we have discarded  the displacement current term $\frac{\partial\bm E}{\partial t}$. 

 In order to simplify the further calculation, we consider only the transverse waves
in that case the Bohm potential i.e.  $<Q_\alpha>$ term in Eqn.(\ref{F_Q}) can be dropped \cite{Haas_11}. Further all the other terms in Eqn.(\ref{F_Q}) are 
are second order in the spin variable and of order $\hbar^2$. We neglect these terms.
However, $\bm F_{\nu e}$ term in Eqn. (\ref{Hydro_eqn_V_q}) and the spin dynamic Eqn. (\ref{Hydro_eqn_S_q}) are order $\hbar$. These
terms are retained in the calculation. In such a situation can write the total force density exerted on the fluid element as,

\begin{equation}
	F^i=-\partial^i\Big(\frac{B_0^2}{2}-\bm B\bm\cdot M\Big)+\bm B\cdot\nabla H^i-\partial^i P-\partial^j \Pi^{ij}
\end{equation}

For an isotropic plasma, the trace-free part of the pressure tensor $\Pi^{ij}$ is zero. It is worth noting that the spacial part of the stress tensor take the form $T^{ij}=-H^iB^j+\big(B^2/2-\bm B\cdot \bm M\big)\delta^{ij}$ \cite{Robinson_00}, apart from the pressure terms. Thus the total force density on a magnetized fluid element can be written as, $F^i=-\partial^j T^{ij}$. Therefore the momentum conservation equation takes the form,

\begin{equation}
	\rho\Big(\frac{\partial}{\partial t}+\bm V\cdot\nabla\Big)\bm V=-\nabla\Big(\frac{B^2}{2}-\bm B\bm\cdot \bm M\Big)+\Big(\bm B\cdot\nabla\Big)\bm H-\nabla P-\nabla\cdot\bm\Pi
	\label{mom_approx_fluid}
\end{equation}

Following the procedure in Ref. \cite{Boyd_69} we can write,

\begin{equation}
	\bm j\sim\frac{\sigma m_i}{\rho e}\nabla P+\sigma\Big(
	\bm E+\bm V\times\bm B\Big)+\frac{\sigma m_i}{\rho e}\bm j\times\bm B+\frac{\sigma m_i}{\rho e}\bm {F}_Q-n_0\frac{\sigma m_i}{\rho}\Big(\frac{\Delta f^0}{2m_e}\Big)\bm S\times\bm B
	\label{j_fluid}
\end{equation}

Taking $\rho\sim n_0 m_i$ the expression for total current can be written as,

\begin{equation}
	\bm j\sim\sigma \Big(\bm E+\bm V\times \bm B\Big)-\sigma\Big(\frac{\Delta f^0}{2m_e}\Big)\bm S\times\bm B+\bm{j}_M
\end{equation}

For the above expression for the hydrodynamic current, the time evolution for the magnetic field $\bm B$ is given by

\begin{equation}
	\frac{\partial\bm B}{\partial t}=-\eta\nabla\times\Big(\nabla\times \bm B\Big)+\nabla\times\Big(\bm V\times\bm B\Big)-\Big(\frac{\Delta f^0}{2m_e}\Big)\nabla\times\Big(\bm S\times\bm B\Big)+\eta\nabla\times\bm{j}_M
	\label{B_with _finite_eta}
\end{equation} 

Where, $\eta=1/\sigma$ is the resistivity.


\section{Neutrino induced vorticity,  Alfv\'en wave and normal modes } 
\label{mode_analysis}

In this section, we consider a very simple scenario. A background magnetic field $\bm B_0=B_0 \bm{\hat z}$ is applied to the plasma. As a result, there is a non-zero constant magnetization in the system even in the absence of any perturbations, which also imply that $\bm S\times \bm B=0$ for the plasma at equilibrium. In this case the spin of the electrons align anti-parallel to the magnetic field to reduce the energy and there fore we can assume the equilibrium magnetization density $\bm M_0$ to take the form $\bm M_0=-\mu_B n_e \bm S_0=\mu_Bn_e\xi\big(\frac{\mu_B B_0}{T_e}\big)\bm{\hat z}$, where $\xi (x)=tanh(x)$ is the Brillouin function. For the following discussions we make the approximation $\xi\big(\frac{\mu_B B_0}{T_e}\big)\sim \big(\frac{\mu_B B_0}{T_e}\big)$ so that $\bm S_0\sim -\frac{1}{2}\frac{\mu_BB_0}{T}\bm{\hat z}$.  Furthermore  we assume that there are  no electromagnetic perturbations in  the system and the fluid velocity enters into the governing equations as perturbation. That is, $\bm E=0, \bm B= B_0\bm{\hat z}, \bm V=\delta\bm V$ and $\bm S=\bm S_0+\bm \delta \bm S$. With these assumptions,  up to linear order  in perturbations, we use the hydrodynamic equations in the following form.

\begin{equation}
	\frac{\partial\delta\bm S}{\partial t}=2\mu_B\delta\bm S\times\bm B_0-\Big(\frac{\Delta f^0}{2m_e}\Big)\bm S_0\times\delta\bm V
\label{spin_evo_no_em}
\end{equation}
\begin{equation}
\rho_0\frac{\partial\delta\bm V}{\partial t}=-\frac{\mu_Bn_e}{T}\Delta f^0\nabla(\bm B_0\cdot\delta\bm V)+
\frac{\mu_Bn_e}{T}\Delta f^0(\bm B_0\cdot\nabla)\delta \bm V	
\label{mom_con_no_em}
\end{equation}


From the equation  (\ref{spin_evo_no_em}) we get  $\frac{\partial\delta\bm S\cdot\bm B_0}{\partial t}=0$. In order to satisfy this conditions, we choose  $\delta\bm S\cdot\bm B_0=0$.  We also take the space-time dependence of the perturbations to be of the following form,
\begin{equation}
	\delta\bm V(t,\bm x)=\delta\bm V_{\omega,\bm k}e^{-i\big(\omega t-\bm k\cdot\bm x\big)},~~~~ \delta\bm S(t,\bm x)=\delta\bm S_{\omega,\bm k}e^{-i\big(\omega t-\bm k\cdot\bm x\big)}
\end{equation}
With these assumptions we get,
\begin{equation}
	\delta\bm S_{\omega,\bm k}=-\Big(\frac{\Delta f^0}{2\Omega_e}\Big)|\bm S_0|\delta\bm V_{\omega,\bm k} 
	\label{spin_no_em}
\end{equation} 
Where, $\Omega_e=\frac{eB_0}{m_e}$. To obtain the above expression we have assumed that $\frac{\omega^2}{\Omega_e^2}\ll 1$. Thus, from the equations (\ref{mom_con_no_em}),  and (\ref{spin_no_em}), we get
\begin{equation}
	-i\omega \delta\bm V_{\omega,\bm k}=\Big(\frac{\mu_Bn_e}{T}\Delta f^0\Big)\bm\Omega_{\omega,\bm k}\times\bm{B}_0
	\label{neu_ind_vor_no_em}
\end{equation}
Where,  $\bm\Omega_{\omega,\bm k}=i\bm k\times\delta\bm V_{\omega,\bm k}$ is the vorticity in the Fourier space. Note that, in the above expressions, we have kept the terms only up to linear in $\Delta f^0$. From the above expression we can see that vorticity term will not contribute to the fluid dynamics if $\Delta f^0=0$. Therefore we conclude that $\bm\Omega$ is induced via the electron neutrino interaction . From the Eqn. (\ref{neu_ind_vor_no_em}) we can obtain the dispersion relations. For the case $\bm k ||\bm B_0$,
\begin{equation}
	\omega=-\Big(\frac{\mu_B n_e B_0}{\rho_0}\Big)\Big(\frac{\Delta f^0}{T}\Big)k
	\label{neu_ind_vor_no_em_w_para}
\end{equation}
 Group velocity of this new mode is given by,
\begin{align} 
	v_g=&\bigg|\frac{d\omega}{dk}\bigg|
	=\bigg |\Big(\frac{\mu_Bn_e B_0}{\rho_0}\Big)\Big(\frac{\Delta f^0}{T}\Big)\bigg| \\
	\sim & 2\sqrt 2\Big(\frac{\mu_Bn_e B_0}{\rho_0}\Big)\Big(\frac{G_F}{T}\Big)\bigg | n_{\nu e}-n_{\bar\nu e}\bigg |
	\label{new_mod_vg}
\end{align}
The  Eqn. (\ref{neu_ind_vor_no_em_w_para}) corresponds to a new type of transverse mode propagating in the direction parallel to the background magnetic field, \textit{induced} by asymmetry in the neutrino background. The wave velocity not just depend on the strength of magnetic field but also on the neutrino asymmetry. This new mode
is similar to the one found in very high energy plasma with the chiral-anomaly \cite{Yamamotto_16}.  In contrast to Ref. \cite{Yamamotto_16}, in our work
the electrons are not considered to be chirally polarized. However, the parity violating
interaction in our work arises due to  neutrino-electron interaction. Further, the effect
of dissipation can easily be introduced by incorporating contribution of the finite shear viscosity  $-ik^2\eta_{vis}$ and the resistivity $-i\sigma_1 B^2_0$ into the dispersion relation (\ref{neu_ind_vor_no_em_w_para}), where $\eta_{vis}$ is the kinematic viscosity and 
$\sigma_1=\sigma/\rho_0$ with $\sigma$ being the resistivity.


Next,  we consider  the effect of  electro-magnetic perturbations. That is we take the perturbations in the following  form.
\begin{equation}
\bm V=\delta\bm V,~~\bm B=B_0\bm{\hat z}+\delta\bm B,~~	\bm E=\delta\bm E.
\end{equation}

For this case, linearized hydrodynamic equations, Eqn.(\ref{cont_flu}) and (\ref{mom_approx_fluid}) takes the form,
\begin{equation}
	\frac{\partial\rho}{\partial t}+\rho_0\nabla\cdot\delta\bm V=0
	\label{cont_with_em}
\end{equation} 
\begin{equation}
\rho_0\frac{\partial\delta\bm V}{\partial t}=-\nabla\Big(\bm {B}_0\cdot\delta\bm B-\bm {M}_0\cdot\delta
\bm B-\delta\bm M\cdot\bm {B}_0\Big)+\bm{B}_0\cdot\nabla\delta\bm H-\nabla P
\label{mom_with_em}	
\end{equation}
And the spin evolution equation becomes,
\begin{equation}
	\Big(\frac{\partial\bm S}{\partial t}\Big)=2\mu_B\bm S\times\bm B-\frac{\Delta f^0}{2}\bm S\times\delta\bm V
\end{equation}
Where, $\bm S=\bm{S}_0+\delta\bm S$. For a perfectly conducting medium $(\eta\rightarrow 0)$, the equation (\ref{B_with _finite_eta}) becomes,
\begin{equation}
	\frac{\partial\delta\bm B}{\partial t}=\nabla\times\Big(\delta\bm V\times\bm {B}_0\Big)-\Big(\frac{\Delta f}{2m_e}\Big)\nabla\times\Big(\delta\bm S\times\bm {B}_0+\bm {S}_0\times\delta\bm B\Big)
	\label{evo_delta_b_with_em}
\end{equation}
Following the same procedure in the last section with same assumptions, we get the expression for $\delta\bm S$ as,
\begin{equation}
	\delta\bm S_{\omega,\bm k}=\frac{\mu_B}{T}\delta\bm B_{\omega,\bm k}-\frac{\Delta f^0}{T}\delta\bm V_{\omega,\bm k}
	\label{spin_with_em}
\end{equation}
Where, $\omega_p^2$ is the plasma frequency. Using the equations (\ref{cont_with_em}), (\ref{mom_with_em}) and (\ref{spin_with_em}) and using the approximation  $\bm M_0=-\mu_B n_e \bm S_0=\mu_Bn_e\eta\big(\frac{\mu_B B_0}{T_e}\big)\bm{\hat z}$ we get,
\begin{equation}
-i\omega\rho_0\delta\bm V_{\omega, \bm k}=i\Big[\frac{\omega_p^2}{m_eT}\Big]\bm {B}_0\times\Big(\bm k\times\delta\bm B_{\bm k,\omega}\Big)-\Big[\frac{\mu_Bn_e}{T}\Delta f^0\Big]\bm {B}_0\times\bm\Omega_{\bm k,\omega}	
\label{evo_delta_V_with_em}
\end{equation}

 Where, $\bm\Omega_{\bm k,\omega}=i\bm k\times\delta\bm V$ is the vorticity in the Fourier space. We can see that the last term in the Eqn.(\ref{evo_delta_V_with_em} ) is proportional to the neutrino asymmetry of the background. expression for velocity in the Fourier space as,
\begin{equation}
	\delta\bm V_{\bm k,\omega}=\Bigg(\frac{\bm B_0\cdot\bm k}{\rho_0\omega^2}\Big[\frac{\omega_p^2}{m_eT}-1 \Big]+\frac{1}{\omega}\frac{\mu_B n_e}{\rho_0 T}\Delta f^0\Bigg)\Bigg((\bm B_0\cdot\delta\bm V_{\omega,\bm k})\bm k-(\bm B_0\bm\cdot \bm k)\delta\bm V_{\omega,\bm k} \Bigg)
\end{equation}
Note that, we have neglected the contributions from the pressure terms in the above expression.  Taking $\bm k$ in the direction of background magnetic field and assuming $\bm {B}_0\cdot \delta\bm V=0$, we get the following dispersion relation,
\begin{equation}
\omega=-\frac{\tilde v_A}{\sqrt{\rho_0\alpha}}\frac{\mu_B n_e}{2T}\Delta f^0 k\pm \tilde v_A k
\label{omega_no_eta_with_em}
\end{equation}

where, $\alpha=(1-\frac{\omega_p^2}{m_eT})$ and $\tilde v_A=v_A\alpha^{1/2}$ is the spin-modified Alfv\'en velocity \cite{Brodin_07}. Here we note that the quantity $\alpha$ describes the spin corrections and in the absence of spin dynamics $\alpha=1$.  It is clear from the Eqn. (\ref{omega_no_eta_with_em}) that the   group velocity $v_g$ can have two values given by,
\begin{equation}
 v_g^{\pm}=\tilde v_A\bigg|1\pm\frac{1}{\sqrt{\rho_0\alpha}}\frac{\mu_B n_e}{2T}\Delta f^0 \bigg|
 \label{group_v}
\end{equation}
which is absent in the absence of any neutrino asymmetry ($\Delta f^0=0)$. Thus we can have two different group velocities for the Alfv\'en waves propagating parallel or anti-parallel to $\bm B_0$.

For finite value of conductivity, we have to take into account of the first and last terms of the Eqn.(\ref{B_with _finite_eta}) and the dispersion relation can be obtained from,
\begin{equation}
	\omega^2+\omega\Big[i\alpha\eta k^2+\frac{\mu_Bn_e\Delta f^0}{T\sqrt{\rho_0\alpha}}\tilde v_Ak\Big]-\tilde v_A^2k^2=0
\end{equation}
Solving for $\omega$ we get,
\begin{equation}
	\omega=-\frac{1}{2}\bigg[i\alpha\eta k^2+\frac{\mu_B n_e \tilde v_A}{T\sqrt{\rho_0\alpha}}\Delta f^0 k\bigg]\pm\frac{1}{2}\sqrt{\bigg[i\alpha\eta k^2+\frac{\mu_B n_e \tilde v_A}{T\sqrt{\rho_0\alpha}}\Delta f^0 k\bigg]^2+4\tilde v_A^2k^2}
	\label{omega_eqn_with_eta}
\end{equation}
 We can see that in the absence of any neutrino asymmetry and $\eta$, the Eqn.(\ref{omega_eqn_with_eta}) reduces to $\omega^2=\tilde v_A^2k^2$, which is the same in magneto hydrodynamics  with spin corrections as obtained in \cite{Brodin_07}.  
\section{Neutrino asymmetry and the  pulsar kick}
\label{pul_kick}
We use our formalism for a qualitative calculation  of observed pulsar kick \cite{Minkowski_1970,Lyne_1982,Hansen_1997}. There are several attempts to explain the reason for the kick, for eg. see the references \cite{Gott_1970,Iben_1996,Kusenko_1999,Kusenko_2004}. Recently there have been attempts to explain the pulsar kick using anomalous hydrodynamic theories (for eg. see Ref.\cite{kaminski2016anomalous}), but the exact reason  for the pulsar kick is  not yet resolved.

   We note that, the energy flux associated with the wave is equal to the  energy density in the wave  times the group velocity \cite{Stix_1992}, which is the Poynting vector $\bm P=\bm E\times\bm B$ in our case \cite{Stix_1992,Cramer_2011}.  The Poynting vector can be expressed in the form,

\begin{equation} 
\bm P=(\omega  A^2)\bm k
\label{Poynting_defi}
\end{equation} 
where $A$ is the magnitude of the vector potential $\bm A_{\omega,\bm k}$. Using the Eqn. (\ref{omega_no_eta_with_em}) we write,

\begin{align}
|\bm P|&=k^2 A^2\tilde v_A\Big(1\pm\frac{1}{\sqrt{\rho_0\alpha}}\frac{\mu_B n_e}{2T}\Delta f^0 \Big)\label{mod_flux}\\
&=\Big(k^2A^2\Big)v_g\label{E_den_times_vg}
\end{align}

From Eqn. (\ref{E_den_times_vg}), we infer that the energy density associated with the wave is $k^2A^2$. Further we note from the Eqn. (\ref{mod_flux}) that the energy transported in the direction of the background field $\bm B_0$ and opposite to $\bm B_0$ are different due to the parity violation within the system. An excess amount of energy is transported in the direction of magnetic field. This excess amount of energy transported per unit area per unit time is given by,

\begin{equation}
\Delta P=\Big(k^2A^2\tilde v_A\Big)\Big(\frac{\Delta f^0}{T}\frac{\mu_Bn_e}{\sqrt{\rho_0\alpha}}\Big)
\label{Delta_P}
\end{equation} 

Which is essentially the  momentum carried by the \textit{excess} photons  leaving the pulsar per unit area per unit time. Therefore the change in velocity experienced by the pulsar can be expressed as,

\begin{equation}
\Delta V_{NS}=\frac{\Delta P}{M_{NS}}\times \Delta t\times (area)
\label{Delta V_NS_defi}
\end{equation}

Where $M_{NS}\sim 10^{30}~kg$ is the mass of neutron star and $\Delta t$ is the time span we assume for the  kick to last, which is approximately $10$ seconds. The radius of the neutron star $R_{NS}$ is approximately  $10~km$. Taking $k\sim A\sim T$,  $\Delta n_{\nu e}\sim 1.6\times 10^{8}~(MeV)^3$,  $T\sim 10^{12}~K $ and $B_0\sim(10^{15}-10^{16})~Gauss$
\cite{Andreas_Reisenegger}, we get $\Delta V_{NS}\sim (10^2-10^3)~km/s$,   which is within the order of  magnitude of observed pulsar kicks. 
 
\section{Discussion and conclusion}	
\label{discussion}

 In conclusion we have developed spin magnetohydrodynamic equations in the presence of asymmetric background neutrinos and  analysed the normal modes of the plasma
in presence of a constant magnetic field. We have shown that a 
new kind of wave (Alfv\'en) is generated Eqn.\ref{neu_ind_vor_no_em_w_para} can exist
whose velocity depends on the neutrino asymmetry. Such a wave can be generated in
a dense astrophysical plasma such as magnetar. For example 
for $B_0^{15}$ Gauss, $T\sim 10 MeV$ and $\Delta n_{\nu e}\sim 1.6\times 10^{8}$(MeV)$^3$, one can estimate the velocity
of the wave (in units of speed of light) around $10^{-5}$. Similarly for the Alfv\'en
waves (Eqn. (\ref{group_v}) ) can have two different speeds. 
We have shown that the background neutrino asymmetry can change the wave-velocity in directions parallel and anti-parallel to the external magnetic field (as shown in Eqn. (\ref{group_v})). We have used our formalism to calculate the kick received by pulsar during its birth. An order of magnitude calculation matches with the observations $\Delta V_{NS}\simeq(10^2-10^3)~km/s$. In the appendix we have 
derived the relativistic hydrodynamical equation for the electrons using Dirac equation. For the case when the electrons
are relativistic the estimate given here for the Alfv\'en velocity can be suppressed by a factor $1/\sqrt{2}$. 

\begin{appendix}
\section{}

	\label{app_A}
In the many astrophysical situations it is necessary to consider the system temperature to be greater
than its rest-mass and therefore we discuss relativistic generalization of  the electron fluid.
For such a generalization in the context of quantum plasma one needs to start with the Dirac equation.
Works by Pauli \cite{WP_36}, Harish Chandra \cite{HC_45} and T. Takabayasi\cite{TT_57} have shown that Dirac equation can be cast into hydrodynamical form. 
Here we use the methodology similar to that given in \cite{TT_57}(see also \cite{FA_11})
to describe the fluid equations for relativistic electrons in presence of the asymmetric neutrino
background. In the standard MHD-approximation electron contributes in defining the current whereas
the ion provides the inertia and therefore significantly contributes to the fluid velocity
[\cite{Boyd_69},\cite{Brodin_07}]. In this appendix we first derive electron fluid equations from
the Dirac equation and then carry-out the MHD approximations with the non-relativistic ion fluid
and obtain an expression for the relativistic corrections to the MHD current and finally
 discuss the changes this brings about the our (non-relativistic) results on the Alfv\'en waves.
The subsequent derivation is rather lengthy and involved we would like to refer the readers to Ref.\cite{TT_57} for further details.

 Following \cite{TT_57} we start with writing the bilinear covariants with hydrodynamic variables and establishing the relations among them using properties of the gamma matrices. 
And also establishing their evolution equations from moments of the corresponding Dirac equation. We choose following bilinear covariants.
\begin{align}
\Omega=&\bar\psi\psi \\
\bar\Omega=&i\bar\psi\gamma^5\psi\\
S^\mu=&\bar\psi\gamma^\mu\psi \\
\bar S^\mu=&\bar\psi\gamma^5\gamma^\mu\psi\\	
M^{\mu\nu}=&\bar\psi\sigma^{\mu\nu}\psi  \\%
\bar M^{\mu\nu}=&i\bar\psi\gamma^5\sigma^{\mu\nu}\psi 
\end{align}
Where, $\sigma^{\mu\nu}=(i/2)[\gamma^\mu, \gamma^\nu]$.
One can obtain the  equations of motion for $\psi$ and $\bar\psi$ from Eqn.(\ref{lagrangian}) and
using these equations of motion one can write following two generic equations involving the dynamics of the  above bilinear forms:
\begin{equation}
	i\bigg(\bar\psi\gamma^A\gamma^\mu\partial_\mu\psi+\partial_\mu\bar\psi\gamma^\mu\gamma^A\psi\bigg)-eA_\mu\bar\psi[\gamma^A, \gamma^\mu]\psi-\frac{\Delta f_\mu}{2}\bar\psi[\gamma^A, \gamma^\mu\gamma^5]\psi=0
	\label{D_Eq+}
	\end{equation}
	\begin{equation}
	i\bigg(\bar\psi\gamma^A\gamma^\mu\partial_\mu\psi-\partial_\mu\bar\psi\gamma^\mu\gamma^A\psi\bigg)-eA_\mu\bar\psi\{\gamma^A, \gamma^\mu\}\psi-\frac{\Delta f_\mu}{2}\bar\psi\{\gamma^A, \gamma^\mu\gamma^5\}\psi-2m\bar\psi\gamma^A\psi=0
	\label{D_Eq-}
	\end{equation}
Using the definition for covariant differential operator $\delta^*_\mu(\bar\psi\gamma^A\psi)=i(\bar\psi\gamma^A\partial_\mu\psi-\partial_\mu\bar\psi\gamma^A\psi)-2eA_\mu\bar\psi\gamma^A\psi $ we define,
\begin{align}
j_\mu=&(1/2m)\delta^*_\mu\Omega \label{Def_jmu}\\
\bar j_\mu=&(1/2m)\delta^*_\mu\bar\Omega \\
T_\mu^\nu=&(1/2m)\delta^*_\mu S^\nu \\
\bar T_\mu^\nu=&(1/2m)\delta^*_\mu \bar S^\nu \\
N^{\mu\nu}_\alpha=&(1/2m)\delta^*_\alpha M^{\mu\nu}\\
\bar N^{\mu\nu}_\alpha=&(1/2m)\delta^*_\alpha \bar M^{\mu\nu}
\end{align}
The quantities $M^{\mu\nu}$ and $\bar M^{\mu\nu}$ can be expressed as $\rho^2 M^{\mu\nu}=-\bar\Omega( S^\mu\bar S^\nu-S^\nu\bar S^\mu)+\Omega\epsilon^{\mu\nu\kappa\lambda}S_\kappa\bar S_\lambda$ and $\bar M^{\mu\nu}=(1/2)\epsilon^{\mu\nu\kappa\lambda} M_{\kappa\lambda}$, where $\rho=\sqrt{\Omega^2+\bar\Omega^2}$ has the interpretation of density.
From Eqns.(\ref{D_Eq+}) and (\ref{D_Eq-}) we obtain the evolution equation of the above defined quantities. 
\begin{align}
\partial_\mu S^\mu=&0 \label{evo_1}\\
\partial_\mu\bar S^\mu =&-2m\bar\Omega \label{evo_2}\\
(1/2m)\partial_\nu M^{\mu\nu} +j^\mu-S^\mu+&\frac{\Delta f_\nu}{2m}\bar M^{\mu\nu}=0 \label{evo_3}\\
(1/2m)\partial_\nu \bar M^{\mu\nu} +\bar j^\mu-&\frac{\Delta f_\nu}{2m} M^{\mu\nu}=0 \label{evo_3}
\end{align}
Next, one defines\cite{FA_11} four-velocity $v_\mu=S_\mu/\rho$ and four-spin $w_\mu={\bar S}_\mu/\rho$ in such a way
that it satisfies the following constraints: $v^\mu v_\mu=1$, $w^\mu w_\mu=-1$ and $v^\mu W_\mu=0$. 
From the last constraint, it is clear that $w_0={\mathbf v}\cdot {\mathbf w}/u^0$ and thus in the
rest-frame zeroth component of the four spin $w_0=0$. 
By taking the divergence of ${\bar T}^{\mu \nu}$  one obtains the following equation
\begin{equation}
	\partial_\nu{\bar T}^{\mu\nu}=\frac{e}{m}w_\nu F^{\mu\nu}-{\bar j}^\mu_{st}+
\delta f_\nu\left[-\rho sin\theta\,\left(v^\mu w^\nu-v^\nu w^\mu\right)+
\rho cos\theta\,\epsilon^{\mu\nu\kappa\lambda}v_\kappa w_\lambda \right]
\label{spin1}
\end{equation}
where we have used $M^{\mu\nu}=\left[-\rho sin\theta\,\left(v^\mu w^\nu-v^\nu w^\mu\right)+
\rho cos\theta\,\epsilon^{\mu\nu\kappa\lambda}v_\kappa w_\lambda \right]$ following \cite{TT_57} and ${\bar j}^{\mu}_{st}$ has the same standard expression as given in Refs.[\cite{TT_57} or \cite{FA_11}]. Besides we have used
new definitions: $cos\theta=\Omega/\rho$ and $sin\theta={{\bar\Omega}/\rho}$. Here we note that $f^{\mu}$ term for the neutrino current does not appear in the above equation. Equation (\ref{spin1}) is at the single body particle-antiparticle state level and one is required to take the fluid average for a collection of $N$ such states. This $N$ particle spinor must be written as a $4^N\times 4^N $ Slatter determinant
of $N$ one-particle states this procedure had been developed in Ref.\cite{FA_11} and we follow it up for our calculation. We find following equation,
in thermal equilibrium, for the spatial part of the spin dynamic:
\begin{equation}
\gamma\left(\partial_t+{\mathbf v}\cdot\nabla\right)	{\mathbf S}=\frac{e/m}{<cos\theta>}
\left(W^0{\mathbf E}/2+ {\mathbf S}\times{\mathbf B}\right)-
\gamma\Delta f^0{\bf S}\times {\mathbf v}+
\gamma\Delta f^0\frac{<sin\theta>}{<cos\theta>}{\mathbf S}
\label{rel_spin_dyn}
\end{equation}
where, $\gamma=\frac{1}{\sqrt{1-v^2}}$ and $W^0={\mathbf S}\cdot {\mathbf v}$. 
Here we note that as we have assumed before we have dropped the spin-thermal coupling and the non-linear spin terms.
The last two terms on the right hand side gives an additional contribution to the spin dynamics of the electron-fluid dynamics given in Ref.\cite{FA_11}. This additional term solely depends on the neutrino background as it should be the case. Following electron relativistic hydrodynamical model in Ref.\cite{FA_11}, we we regard $\theta$ as a constant parameter which is zero for the non-relativistic quantum case and for an extreme relativistic quantum case $\theta=\pi/4$. For the non-relativistic spin dynamics (Eqn.(\ref{Hydro_eqn_S_q})) can be reproduced when we take $\theta=0$. When the electrons are at relativistic temperature one can
replace $m n$ by $(\epsilon+p)$ i.e. by enthalpy density \cite{LLB}. 

 Now one can define the total mass density $\rho=(\epsilon+p)+m_i n_i$ where, $(\epsilon+p)$ represents 
the enthalpy density of the electrons. Since in the magnetohydrodynamic equations  the inertia of
the fluid is dominated by ions, the momentum of the fluid is dominated by the ion momenta \cite{Boyd_69}
(see also \cite{Brodin_07}). This remains true for the relativistic electron case also provided $\rho\sim m_i n_i$ and Eqn.(\ref{mom_approx_fluid}) remains valid for us. 
 Next we derive the analogue of Eqn.(\ref{neu_ind_vor_no_em_w_para}) when the electrons are relativistic. For this consider that there
is an external magnetic field $B_0$ in  $z-$direction and there is no streaming of fluid ${\mathbf v}_0=0$.
Also there is no electromagnetic perturbations i.e. $\delta {\mathbf E} $, $\delta {\mathbf B}$=0. The background spin vector is anti-parallel to the external magnetic field and given by 
$S_0=-\mu_B B_0/2T$ as considered before. Next one can eliminate the electron velocity in the spin
equation ${\mathbf v}_e={\mathbf v}-m_i{\mathbf j}/(Ze\rho)$. Since there is no electromagnetic perturbation
for this case ${\mathbf j}=0$ and one can obtain using Eqs.(\ref{mom_approx_fluid},\ref{rel_spin_dyn}) one obtains the following dispersion 
relation:
\begin{equation}
	\omega=-\left( \frac{\mu_B n_e}{\sqrt{\rho_0}} \right)\left( \frac{\Delta f^0}{2T}<cos\theta>\right)k v_A 
	\label{omega_new_rel_alf}
\end{equation}
where, $v_A=\frac{B_0}{\sqrt{\rho_0}}$. Here we we note here that when 
then the last term in Eqn.(\ref{rel_spin_dyn}) does not contribute to significantly to the dispersion relation.
 
 Similarly for the electromagnetic perturbation for the standard Alfv\'en waves one obtains
the following dispersion relation:
\begin{equation}
	\omega=-\frac{\tilde{v}_A}{\sqrt{\rho_0\alpha}}\frac{\mu_B n_e}{2T}\Delta f^0<cos\theta>k\pm k\tilde{v}_A
	\label{alf_corre_neu+rel}
\end{equation}

Here we note that both the new Alfv\'en waves(Eqn.(\ref{omega_new_rel_alf})) and the regular Alfv\'en waves (Eqn.(\ref{alf_corre_neu+rel})), in the ideal MHD limit, gets
correction due to the relativistic effect which is characterized by
the $<cos\theta>$ factors. Now for an ultra relativistic limit
if one takes $\theta=\pi/4$ following Ref.\cite{FA_11}, one
gets $1/\sqrt{2}$ factor suppression in the speed of the new Alfv\'en wave compared to the non-relativistic case (with 
$\theta=0$) case. 
\end{appendix}



\begin{thebibliography}{unsrt}
%
\bibitem{volpe_14}
C. Volpe,  arXiv:1411.6533.
%
\bibitem{komatsu_10}
E. Komatsu et.al, he Astrophysical Journal Supplement Series. 192 (2): 18 (2010).
%
\bibitem{bashinsky_04}
S. Bashinsky and U. Seljak, Phys. Rev. D {\bf 69}  083002 (2004).
%
\bibitem{kandu_98}
K. Subramaniam and J. D. Barrow, Phys. Rev. D{\bf 58}, 083502 (1998).
%
\bibitem{jedamzik_98}
K. Jademzik, V. Katalinic and A. Olinto, Phys. Rev. D{\bf 57}, 3264 (1997).
%
\bibitem{shaw_10}
J. R. Shaw and A. Lewis, Phys. Rev. D{\bf 81}, 043517 (2010).
%
\bibitem{Bingham_94}
R. Bingham, J.M. Dawson, J. J. Su, H.A. Bethe, 
PhysicsLetters A  
{\textbf 193}, 279 (1994) 
%
\bibitem{Shukla_1997}
P. K. Shukla, L. Stenflo, R.  Bingham,  H. A. Bethe, J. M.  Dawson,  J. T.  Mendon{\c{c}}a, Phys. Lett. A. \textbf{233}, 181 (1997)
\bibitem{Bethe_96}
%
R. Bingham, H.A. Bethe b, J.M. Dawson, P.K. Shukla, J.J. Su, 
Physics Letters A 
 \textbf {220}, 107 (1996) 
 %
\bibitem{Tsytovich_98}
V. N Tsytovich, R. Bingham, J. M Dawson, H. A Bethe,
Astroparticle Physics \textbf {8}, 297 (1998)
%
%
\bibitem{Silva_99}
  Silva, LO and Bingham, R and Dawson, JM and Mori, WB,
  Phys. Rev. E 
  \textbf{59}, {2273} (1999).
  %
\bibitem{Bento_99}
 B. Lu{\'\i}s,
{Phys. Rev. D} 
\textbf{61}, {013004}
 (1999)
%
\bibitem{Esposito_99}
  {Esposito, Salvatore},
{Mod. Phys. Lett. A} 
\textbf{14}, {1763}, (1999).
%
\bibitem{Brizard_00}
{Brizard, Alain J and Murayama, Hitoshi and Wurtele, Jonathan S},
{Phys. Rev. E} 
\textbf{61}, {4410},
 {2000}.
%
\bibitem{Bento_01}
{B. Lu{\'\i}s},
{Phys. Rev. D} 
\textbf{63}, {077302} {2001}.
%
\bibitem{Serbeto_02}
{A. Serbeto and J. T. Mendon{\c{c}}a and P. K. Shukla and L.O. Silva },
{Phys. Lett. A} 
\textbf{305}, {190} (2002)
%
\bibitem{Serbeto_fluid_02}
{A Serbeto },
{Phys. Lett. A} 
\textbf{296}, {217} (2002)
%
\bibitem{Silva_1999}
L. O. Silva, R. Bingham, J. M Dawson and J. T. Mendonc\c{c}a, Phys. Rev. Lett. \textbf{83}, 2703 (1999)
%
\bibitem{Silva_2006}
L. O. Silva and R. Bingham, JCAP \textbf{05}, 011 (2006)
%
\bibitem{Bento_99_recent}
L. Bento, arXiv:hep-ph/9912533v1
%
\bibitem{Palash_89}
P. B. Pal and J. F.Nieves, 
Phys. Rev. D \textbf{39}, 652 (1989)
%
\bibitem{Nieves_00}
J. F. Nieves,
Phys. Rev. D \textbf{61}, 113008 (2000)
%
\bibitem{Nieves_05}
J. F. Nieves and S. Sahu, 
Phys. Rev. D \textbf{71}, 073006 (2015)
%
\bibitem{Boyarsky_12}
 A. Boyarsky,O. Ruchayskiy and M. Shaposhnikov,
 Phys. Rev. Lett. \textbf{109}, 111602 (2012).
 %
\bibitem{Doglov_02}
A. D. Dolgov and D. Grasso,
Phys. Rev. Lett \textbf{88}, 011301 (2002)
%
\bibitem{Vainshtein_72}
S. I. Vainshtein and Ya. B. Zeldovich,
 Usp. Fiz. Nauk \textbf{106}, 431 (1972).
 %
\bibitem{Giunti_07}
C. Giunti and C. W. Kim,
 {\it Fundamentals of Neutrino Physics and Astrophysics}, Oxford University Press, Oxford U.S.A., pg. 138 (2007).
 %
\bibitem{Dornikov_15}
M. Dvornikov and V. B. Semikoz,
	JCAP, \textbf{05}, 002 (2015).
%
\bibitem{Bhatt_16}	
Jitesh R. Bhatt, Manu George, 
{arXiv:1602.06884}.
%

%
\bibitem{Son_13}
D. T. Son  and  N. Yamamoto,
Phys.Rev.D \textbf{87}, 085016 (2013)
%
\bibitem{Son_09}
 D. T. Son and  P. Surowka,
 Phys. Rev. Lett. \textbf{103}, 191601 (2009)
%
\bibitem{Yamamotto_16}
N. Yamamoto, Phys. Rev. D \textbf{93}, 065017 (2016)
%
\bibitem{Dvornikov__galvano_rotational_15}
M. Dvornikov,
JCAP \textbf{05}, 037 (2015)
%
\bibitem{Suvorov_16}
A. G. Suvorov, A. Mastrano and A. Melatos, Monthly Notices of the Royal Astronomical Society \textbf{456}, 731 (2016)
%
\bibitem{Diaz_16}
J. S. D{\'{i}}az and F. R. Klinkhamer,
Phys. Rev. D \textbf{93}, 053004 (2016)
%
\bibitem{Brodin_07}
G. Brodin and M. Marklund,
 New Journal of Physics
 \textbf {9} (2007) 277 
 %
 \bibitem{Mahajan_11}
 S. M. Mahajan . and F. A. Asenjo,
 Phys. Rev. Lett.\textbf{107}, 195003, 
 2011
 %
 \bibitem{Greiner_reqm}
W. Greiner, 
Relativistic quantum mechanics. Vol. \textbf{3}. Berlin: Spinger (1990)
 %
\bibitem{Robinson_00}
M. P. Robinson  et al.
 Phys. Rev. Lett. 
 \textbf {85}, 4466  (2000)
 %
\bibitem{Boyd_69}
T. J. M Boyd and J. J Anderson, "Plasma dynamics",
 Thomas Nelson and sons Ltd. (1969).
%
\bibitem{Kulsrud_82}
R. M. Kulsrud, H. P. Furth, E. J. Valeo and M. Goldhaber, Phys.Rev. Lett. {\bf 49}, 1248 
(1982).
%
\bibitem{Cowley_86}
S. C. Cowley, R. M. Kulsrud and E. Valeo, Phys. Fluids {\bf 49},430 (1986).
%
\bibitem{Walser_02}
M. W. Walser, D. J. Urbach, K. Z. Hatsagortsyan, S. Hu and C.H. Keitel,
Phys. Rev. {\bf A 65}, 043410 (2002).
%
\bibitem{Haas_11}
F. Haas, {\it  Quantum plasmas: An hydrodynamic approach},
Sringer, (2011).
%
\bibitem{Minkowski_1970}
R. Minkowski, PASP. \textbf{82}, 470 (1970) 
%

\bibitem{Lyne_1982}
A. G. Lyne, B. Anderson and M. J Salter, MNRAS. \textbf{291}, 503, (1982).
%
\bibitem{Hansen_1997}
B. M. S. Hansen and  E. S. Phinney, MNRAS. \textbf{291}, 569,(1997). 
%
\bibitem{Gott_1970}
J. R. Gott, J. E. Gunn and J. P. Ostriker, ApJ. \textbf{160}, L91, (1970).
%
\bibitem{Iben_1996}
I. Iben, A. V. Tutukov, ApJ. \textbf{456}, 738 (1996).
%
\bibitem{Kusenko_1999}
A. Kusenko, G. Segr\'e, Phys. Rev. D, \textbf{59}, 061302 (1999).
%
\bibitem{Kusenko_2004}
A. Kusenko, Int.J.Mod.Phys. D \textbf{13}, 2065 (2004). 
%
\bibitem{kaminski2016anomalous}
Matthias kaminiski, Christoph F. Uhlemann, Marcus Bleicher and J{\"u}rgen Schaffner-Bielich, Phys. Lett.B \textbf{760},170 (2016). 
%
\bibitem{Stix_1992}
T. H. Stix, \textit{Waves in plasmas}, Springer Science \& Business Media (1992).
%
\bibitem{Cramer_2011}
Niel. F. Cramer, \textit{The physics of Alfv{\'e}n waves}, John Wiley \& Sons (2011). 
%
\bibitem{Andreas_Reisenegger}
Andreas Reisenegger, 	arXiv:1305.2542
\bibitem{WP_36}
W. Pauli,
Ann. Inst. Poincar{\'e} \textbf{6}, 109 (1936).
%
\bibitem{HC_45}
Harish-Chandra,
Proc. Indian Acad. Sci. (Math. Sci.) \textbf{22}, 30 (1945).

\bibitem{TT_57}
T. Takabayasi,
Progress of Theoretical Physics Supplement \textbf{4}, 1 (1957).
\bibitem{FA_11}
F. A. Asenjo, V. Mu{n}oz, J.A. Valdivia and S. M. Mahajan,
Physics of plasmas \textbf{18}, 012107 (2011).
%
\bibitem{LLB} L.D. Landau and E. M. Lifshitz, 
{\it Fluid Mehcanics}, Pergamon Press, 2nd edition (1989).

\end{thebibliography}
\end{document}